\documentclass[12pt]{article}
\usepackage{amsfonts,amssymb}
\makeatletter 
\def\EquationsBySection{\def\theequation
{\thesection.\arabic{equation}}%
\@addtoreset{equation}{section}}

\newcommand\old[1]{}
\newtheorem{theorem}{Theorem}[section]
\newtheorem{definition}[theorem]{Definition}
\newtheorem{lemma}[theorem]{Lemma}

\newtheorem{example}[theorem]{Ex}

\EquationsBySection \makeatother

\begin{document}
\pagestyle{plain}
\title
{\bf Degree-distribution Stability of Growing
Networks
\thanks{Supported by National Natural Science Foundation of
China (No.10671212). }}
\author{Zhenting Hou$^{1}$\thanks{Email:
zthou@csu.edu.cn}, Xiangxing Kong$^{1}$, Dinghua Shi$^{2}$, Guanrong
Chen$^{3}$,\\ Qinggui Zhao$^{1}$\\
$^{1}$ School of Mathematics, Central South University,\\
 Changsha, Hunan, 410075, China\\
$^{2}$ Department of Mathematics, Shanghai University,\\
 Shanghai 200444, China\\
$^{3}$ Department of Electronic Engineering, City University of Hong Kong,\\
 Hong Kong, China
}
\date{}
\maketitle

{\bf Abstract:}In this paper, we abstract a kind of stochastic
processes from evolving processes of growing networks, this process
is called growing network Markov chains. Thus the existence and the
formulas of degree distribution are transformed to the corresponding
problems of growing network Markov chains. First we investigate the
growing network Markov chains, and obtain the condition in which the
steady degree distribution exists and get its exact formulas. Then
we apply it to various growing networks. With this method, we get a
rigorous, exact and unified solution of the steady degree
distribution for growing networks.

{\bf Key Words:} Growing network Markov chains; BA model;
Scale-free; Degree distribution. \noindent

{\bf PACS numbers:} 89.75.Da, 87.23.Ge, 89.75.Hc

\section{Introduction}
Barab\'asi and Albert$^{\cite{ba}}$found that for many real-world
networks, e.g., the World Wide Web (WWW), the fraction of vertices
with degree $k$ is proportional over a large range to a power-law
tail, i.e. $P(k)\sim k^{-\gamma}$, where $\gamma$ is a constant
independent of the size of the network. For purpose of opening up
mechanism producing scale-free property, they proposed the well
known BA model and summarized the reasons: growth and preferential
attachment. The proposing of BA model led to a great echo among
people, with hundreds of advanced network models proposed and
studied, it also gave rise to a new upsurge in studying complex
networks. Some important examples are as follows:

\begin{example}[{{\rm BA model$^{\cite{ba}}$}}]
Barab\'asi and Albert et al. proposed a model which starts with a
small number ($m_0$) of vertices, at each time step add a new vertex
with $m$ ($\le m_0$) edges that link the new vertex to $m$ different
vertices already present in the system. To incorporate preferential
attachment, they assumed the probability $\Pi$ that the new vertex
will be connected to a vertex $i$ depends on the connectivity $k_i$
of vertex $i$, that
 is $\Pi(k_i)=mk_i/\sum_j k_j$. After $t$ steps the model leads to a
random network with $t+m_0$ vertices and $mt$ edges. Let $k_i(t)$
denotes the degree of vertex $i$ at time $t$, from the mechanism of
the model we know that $k_i(t)$ is a Markov chain.
\end{example}

\begin{example}[{{\rm Growing network with random links}}] Each
vertex is chosen randomly in this case, that is, $\Pi_i=1/t$,
everything else remains the same as BA model.
\end{example}

\begin{example}[{{\rm LL model(i)$^{\cite{l}}$}}]
The model is proposed like this: at each time step a new node with
$m$ links (edges) is added, and the probability $\Pi_i$ is
determined by $m\frac{(1-p)k_i+p}{\sum_j[(1-p)k_j+p]}$, where $0\leq
p\leq1$ is a parameter characterizing the relative weights between
the deterministic and random contributions to $\Pi_i$, and the
summation is over the whole network at a given time. The model
reduces to the BA model for $p=0$ and it becomes a completely random
network for $p=1$. So far the evolving mechanism of the LL model has
not found, so we introduce the following example in which the
boundary between preferential and random are clear.
\end{example}

\begin{example}[{{\rm LL model(ii)}}]
Everything else remains the same as the former example except that
$\Pi_i=\frac{m(1-p)}{2mt+N_{0}}k+\frac{mp}{t+m_0}$.
\end{example}

\begin{example}[{{\rm Generalized collaboration networks$^{\cite{zhzsz}}$}}]
Zhang P.P.,  et al. supposed the initial network to be several
complete graphs with $m_0$ vertices, the sum of degree $k_{i0}$ is
$k_0$. At each time step add a new vertex to the network and connect
it to $T-1$($T$ is a constant) different vertices already present in
the system. The probability that vertex $i$ get a link is
proportional to the degree of the vertex, i.e.
$\Pi_i=k_i/\sum_lk_j$. After this, link all the $T$ vertices to form
a complete graph.
\end{example}

\begin{example}[{{\rm ZRZ model$^{\cite{zrz}}$}}]
The initial network ($t=0$) is a complete graph ($m$-complete graph)
with $m$ vertices and $C_m^2$ edges. At each time step a new vertex
is added to the network, it will connect to all the vertices of a
$m-$complete graph selected randomly, that is $m+1$ people are
collaborated in an act.
\end{example}

\begin{example}[{{\rm KK model$^{\cite{kk}}$}}] At the
beginning ($t=1$) we have one group with one element in it. At each
time step we add a new element to the system. With probability $p$
it will belong to one of existing groups. The probability that it
joins the $i$th group is proportional to the size of the group
$(k_i/N)$ (the number of elements is equal to the time i.e. $N=t$).
With probability $q=1-p$, the new element will belong to a new
group.
\end{example}

\begin{example}[{{\rm DMS model$^{\cite{dms}}$}}] At each time step a new site appears.
Simultaneously, $m$ new directed links coming out from non-specified
sites are introduced. Let the connectivity $q_s$ be the number of
incoming links to a site $s$, i.e., to a site added at time $s$. The
probability that a new link points to a given site $s$ is
proportional to the following characteristic of the site:
$H_s=H+q_s$, thereafter called its attractiveness. All sites are
born with some initial attractiveness $H\geq0$, but afterwards it
increases because of the $q_s$ term. Note that one may allow
multiple links, i.e., the connectivity of a given site may increase
simultaneously by more than one.
\end{example}

\begin{example}[{{\rm LCD model$^{\cite{b}}$}}] Based on BA model, Bollob\'{a}s et al.
proposed another model which allows multiple links and loops.
\end{example}

Except LCD model, the existence and deduction of degree distribution
of the present models (including the noted BA model) are devoid of
exact mathematics basis. Recently, a mechanism for BA model is given
in \cite{dg} and \cite{hksc}. Moreover the existence and exact
formulas of the degree distribution were also provided. In this
paper we abstract a kind of Markov chains from enumerated models, we
call it as growing network Markov chains. First we investigate the
growing network Markov chains, and obtain the condition in which the
steady degree distribution exists and get its exact formulas. Then
we apply it to various growing networks. With this method, we get a
rigorous, exact and unified solution of the steady degree
distribution for growing networks.

\section{Non-multiple Growing Network Markov Chains}
For any $i=1,2,\cdots,k_i(t)(t=i, i+1, \cdots)$ are Markov chains
non-decrease with respect to $t$ taking values in
$\{0,1,2,\cdots\}$. Suppose there exists a positive integer $i_0$,
s.t, $\{k_i(t)\} (i\geq i_0)$ have the initial distribution
$P\{k_i(i)=k\}=d_{k,i}$, with transition probability
\begin{eqnarray}
P\{k_i(t+1)=l|k_i(t)=k\}=\left\{\begin{array}{ll}
f_t(k), &\textrm{$l=k+1,$}\\
1-f_t(k), &\textrm{$l=k,$}\\
0, &\textrm{otherwise.}
\end{array}\right.
\end{eqnarray}
where $0<f_t(k)<1$. Let $P(k,i,t):=P\{k_i(t)=k\} (t=i, i+1, \cdots),
P(k,t):=\frac{1}{t}\sum\limits_{i=1}^tP(k,i,t).$

\begin{definition}
The above Markov chains $\{k_i(t)\}_{t=i,i+1,\cdots}(i=1,2,\cdots)$
are called series of non-multiple growing network Markov chains, for
short we call it non-multiple growing network Markov chains, if the
limit $P(k)=\lim\limits_{t\rightarrow\infty}P(k,t)$ exists, and
\begin{eqnarray}
P(k)\geq0,\sum\limits_{k=0}^\infty P(k)=1.
\end{eqnarray}
we say that the degree distribution of non-multiple growing network
Markov chains exists, and $P(k)$ is the steady degree distribution
of $\{k_i(t)\}$. Further, if $P(k)$ is power-law, i.e.,
\begin{eqnarray}
P(k)\sim k^{-\gamma}(k\geq k_0),
\end{eqnarray}
$\{k_i(t)\}$ are called scale-free non-multiple growing network
Markov chains.
\end{definition}

\begin{lemma}\label{le:1}
If $\lim\limits_{i\rightarrow\infty}d_{k,i}=d_k$ exists and
satisfies $\sum\limits_{k=0}^\infty d_k=1$.
$\lim\limits_{t\rightarrow\infty}tf_t(k):= F(k)$ also exists, and
there is a non-negative integer $m$ to satisfy $d_k=0,
k=0,1,\cdots,m-1, d_m>0$, and $F(k)>0(k=m,m+1,\cdots)$. Then
$\lim\limits_{t\rightarrow\infty}P(m,t)$ exists,  moreover
\begin{eqnarray}\label{le:1.1}
P(m)=\frac{d_{m}}{1+F(m)}.
\end{eqnarray}
\end{lemma}
{\bf Proof} With the Markovian property, we have
\begin{eqnarray}
P(m,i,t+1)=P(m,i,t)[1-f_{t}(m)],\ \ (i\leq t).
\end{eqnarray}

By the definition of $P(m,t)$ and $P(m,t+1,t+1)=d_{m,t+1}$, we
obtain
\begin{eqnarray}
P(m,t+1)=\frac{t}{t+1}P(m,t)[1-f_{t}(m)]+\frac{1}{t+1}d_{m,t+1}.
\end{eqnarray}

The above difference equation has the following solution

\begin{eqnarray}
&
&P(m,t)=\frac{1}{t}\prod\limits_{i=1}^{t-1}[1-f_{i}(m)]\left\{P(m,1)+\sum\limits_{l=1}^{t-1}d_{m,l+1}\prod\limits_{j=1}^{l}[1-f_{j}(m)]^{-1}\right\}.
\end{eqnarray}

Let
$$x_t=P(m,1)+\sum\limits_{l=1}^{t-1}d_{m,l+1}\prod\limits_{j=1}^{l}[1-f_{j}(m)]^{-1},$$
$$y_t=t\prod\limits_{i=1}^{t-1}[1-f_{i}(m)]^{-1}>t\rightarrow\infty.$$

We easily have

$$x_{t+1}-x_t=d_{m,t+1}\prod\limits_{j=1}^{t}[1-f_{j}(m)]^{-1},$$
$$y_{t+1}-y_t=[1+tf_{t}(m)]\prod\limits_{j=1}^{t}[1-f_{j}(m)]^{-1}>0.$$

With $\lim\limits_{t\rightarrow\infty}tf_t(m)=F(m)$, we have
\begin{eqnarray}
\frac{x_{t+1}-x_t}{y_{t+1}-y_t}=\frac{d_{m,t+1}}{1+tf_{t}(m)}\rightarrow\frac{d_{m}}{1+F(m)},\
\ (t\rightarrow\infty).
\end{eqnarray}

With the Stolz-Ces\'{a}ro theorem $^{\cite{s}}$, we have Eq
(\ref{le:1.1}) and complete the proof.

\begin{lemma}\label{le:2}
If the conditions in Lemma \ref{le:1} are all satisfied, and for
$k>m$, $\lim\limits_{t\rightarrow\infty}P(k-1,t)$ exists, then
$\lim\limits_{t\rightarrow\infty}P(k,t)$ exists, moreover
\begin{eqnarray}\label{le:2.1}
P(k)=\frac{F(k-1)}{1+F(k)}P(k-1)+\frac{d_{k}}{1+F(k)}.
\end{eqnarray}
\end{lemma}
{\bf Proof} With the Markovian property, we have
\begin{eqnarray}
P(k,i,t+1)=P(k,i,t)[1-f_{t}(k)]+P(k-1,i,t)f_{t}(k-1)+d_{k,t+1}.
\end{eqnarray}

The definition of $P(k,t)$ and $P(k,t+1,t+1)=d_{k,t+1}$ yield
\begin{eqnarray}
P(k,t+1)=\frac{t}{t+1}P(k,t)[1-f_{t}(k)]+\frac{t}{t+1}P(k-1,t)f_{t}(k-1)+\frac{d_{k,t+1}}{t+1}.
\end{eqnarray}

The above difference equation has the following solution

\begin{eqnarray}
& &P(k,t)=\frac{1}{t}\prod\limits_{i=1}^{t-1}[1-f_{i}(k)]\times\nonumber\\
&
&\{P(k,1)+\sum\limits_{l=1}^{t-1}\left[lP(k-1,l)f_{l}(k-1)+d_{k,l+1}\right]\prod\limits_{j=1}^{l}[1-f_{j}(k)]^{-1}\}.
\end{eqnarray}

Similar to Lemma \ref{le:1} we have Eq(\ref{le:2.1}), then complete
the proof.

\begin{theorem}\label{th:1}
If the conditions in Lemma \ref{le:1} are all satisfied, the steady
degree distribution of $\{k_i(t)\}$ exists, moreover
\begin{eqnarray}\label{th:1:1}
P(k)=\left\{\begin{array}{ll}
\frac{d_{m}}{1+F(m)}, &\textrm{$k=m,$}\\
\prod\limits_{i=m}^{k-1}\frac{F(i)}{1+F(i+1)}[\frac{d_{m}}{1+F(m)}+\sum\limits_{l=m}^{k-1}\frac{\frac{d_{l+1}}{1+F(l+1)}}{\prod\limits_{j=m}^{l}\frac{F(j)}{1+F(j+1)}}], &\textrm{$k>m.$}\\
\end{array}\right.
\end{eqnarray}
\end{theorem}
{\bf Proof}From Lemma \ref{le:1} and Lemma \ref{le:2}, Eq
(\ref{th:1:1}) comes into existence.
\begin{theorem}\label{th:2}
Suppose there is a non-negative integer $M\geq m$ and satisfies
$d_k=0 (k>M)$. And if there two constants $A,B$, satisfy
$F(k)=Ak+B$, then

(I)The degree distribution $P(k)$ satisfies
\begin{eqnarray}\label{th:2.1}
\sum\limits_{k=m}^\infty P(k)=1.
\end{eqnarray}

(II) If $A>0$, then $\{k_i(t)\}$ are Scale-free growing network
Markov chains, and
\begin{eqnarray}
P(k)=\left\{\begin{array}{lll}
\frac{d_{m}}{1+Am+B}, &\textrm{$k=m,$}\\
\prod\limits_{i=m}^{k-1}\frac{Ai+B}{1+A(i+1)+B}[\frac{d_{m}}{1+Am+B}+\sum\limits_{l=m}^{k-1}\frac{\frac{d_{l+1}}{1+A(l+1)+B}}{\prod\limits_{j=m}^{l}\frac{Aj+B}{1+A(j+1)+B}}], &\textrm{$m<k\leq M,$}\\
\frac{\Gamma(k+\frac{B}{A})}{\Gamma(k+\frac{B}{A}+1+\frac{1}{A})}\frac{\Gamma(M+\frac{B}{A}+1+\frac{1}{A})}{\Gamma(M+\frac{B}{A})}\prod\limits_{i=m}^{M-1}\frac{Ai+B}{1+A(i+1)+B}[\frac{d_{m}}{1+Am+B}\\
+\sum\limits_{l=m}^{M-1}\frac{\frac{d_{l+1}}{1+A(l+1)+B}}{\prod\limits_{j=m}^{l}\frac{Aj+B}{1+A(j+1)+B}}]\sim k^{-(1+\frac{1}{A})}, &\textrm{$k>M$.}\\
\end{array}\right.
\end{eqnarray}

Specially, if $d_m=1,d_k=0(k\neq m)$, then
\begin{eqnarray}
P(k)=\left\{\begin{array}{ll}
\frac{1}{1+Am+B}, &\textrm{$k=m,$}\\
\frac{\Gamma(k+\frac{B}{A})}{\Gamma(k+\frac{B}{A}+1+\frac{1}{A})}\frac{\Gamma(m+\frac{B}{A}+1+\frac{1}{A})}{\Gamma(m+\frac{B}{A})}\frac{1}{1+Am+B}\sim
k^{-(1+\frac{1}{A})}, &\textrm{$k>m.$}
\end{array}\right.
\end{eqnarray}

(III) If $A=0,B>0$, $\{k_i(t)\}$ are scale-free growing network
Markov chains.

(IV) The case for $A<0$ or $A=0,B<0$ will never happen.
\end{theorem}
{\bf Proof} From Eq(\ref{le:1.1}), (\ref{le:2.1}) and the given
condition, we easily obtain Eq(\ref{th:2.1}). (II),(III) and (IV)
are also easily proved.

Let $k_i(t)$ denote the degree of the vertex added at time-step $i$
evolved at time $t$ in the former examples, and $\{k_i(t)\}$ are
growing network Markov chains. Therefore we can apply Theorem
\ref{th:1} and Theorem \ref{th:2} to the former examples.

\paragraph{Example 1.1 } We have $d_m=1$ and $f_t(k)=\frac{mk}{2t},$ so
$A=\frac{m}{2},B=0$
\begin{eqnarray}
P(k)=\left\{\begin{array}{ll}
\frac{2}{m+2}, &\textrm{$k=m,$}\\
\frac{\Gamma(k)}{\Gamma(k+3)}\frac{\Gamma(m+3)}{\Gamma(m)}\frac{2}{m+2}\sim 2m^2k^{-3}, &\textrm{$k>m.$}\\
\end{array}\right.
\end{eqnarray}
the network is scale-free with scaling exponent $\gamma=3$. This is
identical with the results in
\cite{ba}\cite{dms}\cite{krl}\cite{dg}\cite{hksc}. A different and
exact proof of degree distribution and an evolving mechanism of this
model have been provided in papers \cite{dg} and \cite{hksc} .

\paragraph{Example 1.2} From the model we
have $d_m=1$ and $f_t(k)=\frac{m}{t},$ so $A=0,B=m$. The steady
degree distribution exists, and
\begin{eqnarray}
P(k)=\frac{m}{1+m}P(k-1)=\left(\frac{m}{1+m}\right)^{k-m}\frac{1}{1+m}.
\end{eqnarray} is exponentially
distributed, the network is not scale-free.

\paragraph{Example 1.3} $d_m=1$ and
$f_t(k)=m\frac{1-p}{(1-p)2mt+pt}k+m\frac{p}{(1-p)2mt+pt}$,
Ôò$A=m\frac{1-p}{(1-p)2m+p}$, $B=m\frac{p}{(1-p)2m+p}$. If $p\neq1$,
the degree distribution
\begin{eqnarray}
P(k)=\left\{\begin{array}{ll}
\frac{2m+(1-2m)p}{m^2+2m+(1-m^2-m)p}, &\textrm{$k=m,$}\\
\frac{\Gamma(k+\frac{p}{1-p})}{\Gamma(k+\frac{p}{1-p}+3+\frac{p}{m(1-p)})}\frac{\Gamma(m+\frac{p}{1-p}+3+\frac{p}{m(1-p)})}{\Gamma(m+\frac{p}{1-p})}\frac{2m+(1-2m)p}{m^2+2m+(1-m^2-m)p}\\
\sim k^{-(3+\frac{p}{m(1-p)})}, &\textrm{$k>m.$}\\
\end{array}\right.
\end{eqnarray}
is power-law with scaling exponent $\gamma=3+\frac{p}{m(1-p)}$.
$P(k)$ follows exponential distribution if $p=1$.

\paragraph{Example 1.4} We have
$d_m=1$,$f_t(k)=m\frac{1-p}{2mt+N_{0}}k+\frac{mp}{t+m_0}.
A=\frac{1-p}{2},B=mp$.
\begin{eqnarray}
P(k)=\left\{\begin{array}{ll}
\frac{2}{2+m+mp}, &\textrm{$k=m,$}\\
\frac{\Gamma(k+\frac{2mp}{1-p})}{\Gamma(k+\frac{2mp}{1-p}+1+\frac{2}{1-p})}\frac{\Gamma(m+\frac{2mp}{1-p}+1+\frac{2}{1-p})}{\Gamma(m+\frac{2mp}{1-p})}\frac{2}{2+m+mp}, &\textrm{$k>m.$}\\
\end{array}\right.
\end{eqnarray}

If $p=1$, $P(k)$ is not power-law, and if $p\neq1$
\begin{eqnarray}
P(k)\sim k^{-(1+\frac{2}{1-p})}
\end{eqnarray} is power-law with the scaling
exponent $\gamma=1+\frac{2}{1-p}$.

\paragraph{Example 1.5}  We get
$d_1=1$,$f_t(k)=\frac{(T-1)k}{k_0+Tt}$, so $A=\frac{T-1}{T},B=0$. So
the network is scale-free and the degree distribution is
\begin{eqnarray}
P(k)\sim k^{-(1+\frac{T}{T-1})}.
\end{eqnarray}
and scaling exponent is $\gamma=1+\frac{T}{T-1}$.

\paragraph{Example 1.6}  $d_m=1(m>2)$,
$f_t(k)=\frac{(m-1)k}{mt+1}-\frac{m(m-2)}{mt+1}$, so
$A=\frac{m-1}{m}, B=-m(m-2)$. The network is scale-free and the
degree distribution is
\begin{eqnarray}
P(k)\sim k^{-(1+\frac{m}{m-1})}.
\end{eqnarray}
 with scaling exponent
$\gamma=1+\frac{m}{m-1}$.

\paragraph{Example 1.7}  $d_0=p,d_1=1-p$, $f_t(k)=p\frac{k}{t}$, so
$A=p$, $B=0$, in this model $k_i(t)$ denotes the number of elements
in group  added at time-step $i$ evolved at time $t$.

The degree distribution is
\begin{eqnarray}
P(k)=\left\{\begin{array}{ll}
p, &\textrm{$k=0,$}\\
\frac{1-p}{1+p}, &\textrm{$k=1,$}\\
\frac{\Gamma(k)\Gamma(2+\frac{1}{p})}{\Gamma(k+1+\frac{1}{p})}\frac{1-p}{1+p},
&\textrm{$k>1.$}
\end{array}\right.
\end{eqnarray}
so it's power-law and the scaling exponent is
$\gamma=1+\frac{1}{p}$.
\section{Multiple Growing Network Markov Chains}
The degree of a vertex can increase at most 1 in network of
non-multiple links. However, the degree can increase more than 1 if
multiple links is permitted. We have found that the probability of
vertex's degree increase more than 1 is the high-level of
infinitesimal. For the purpose of investigating the degree
distribution of multiple linking growing network, we introduce
multiple growing network Markov chains.

For any $i=1,2,\cdots,k_i(t)(t=i, i+1, \cdots)$ are Markov chains
non-decrease with respect to $t$ taking values in
$\{0,1,2,\cdots\}$, suppose there exists a positive integer $i_0$,
s.t, $\{k_i(t)\} (i\geq i_0)$ have the initial distribution
$P\{k_i(i)=k\}=d_{k,i}$, with transient probability matrix

\begin{eqnarray}
   \left(
   \begin{array}{ccccccccc}
   p_{0,0}& p_{0,1}  &o_{0,2}(\frac{1}{t})&\cdots               & o_{0,m'}(\frac{1}{t})           &                 &       &    &  \\
         & p_{1,1}   &p_{1,2}        &o_{1,3}(\frac{1}{t})       &\cdots      & o_{1,m'+1}(\frac{1}{t})                 &       &0  &  \\
         &           &\ddots        &\ddots                 &\ddots      &  \ddots        &\ddots    &  &  \\
         &           &              & p_{K,K} &p_{K,K+1} & o_{K,K+2}(\frac{1}{t})  &\cdots &o_{K,K+m'}(\frac{1}{t}) &   \\
         &    0       &              &                       &1           &0                &\cdots &  &  \\
         &           &              &                       &            &\ddots           &\ddots      &  &  \\
   \end{array}
   \right).
   \end{eqnarray}
where $p_{k,l}=P\{k_i(t+1)=l|k_i(t)=k\}$. For $k\leq K$($K$ is the
maximum degree of vertex $i$ at $t$)
\begin{eqnarray}
p_{k,l}=\left\{\begin{array}{ll}
1-f_t(k)-\sum_{s=k+2}^{k+m'} o_{k,s}(\frac{1}{t}),&\textrm{$l=k,$}\\
f_t(k), &\textrm{$l=k+1,$}\\
o_{k,l}(\frac{1}{t}), &\textrm{$k+1< l\leq k+m',$} \\
0, &\textrm{else.} \\
\end{array}\right.
\end{eqnarray}
and for $k>K$
\begin{eqnarray}
 p_{k,l}=\left\{\begin{array}{ll}
1,&\textrm{$l=k,$}\\
0, &\textrm{$l\neq k+1.$}
\end{array}\right.
\end{eqnarray}

\begin{definition}
The above Markov chains $\{k_i(t)\}_{t=i,i+1,\cdots}(i=1,2,\cdots)$
are called series of multiple growing network Markov chains, for
short we call it multiple growing network Markov chains , if the
limit $P(k)=\lim\limits_{t\rightarrow\infty}P(k,t)$ exists, and
\begin{eqnarray}
P(k)\geq0,\sum\limits_{k=0}^\infty P(k)=1,
\end{eqnarray} we say
that degree distribution of multiple growing network Markov chains
exists, and $P(k)$ is the steady degree distribution of
$\{k_i(t)\}$. Further, if $P(k)$ is power-law, i.e.,
\begin{eqnarray}
P(k)\sim k^{-\gamma}(k\geq k_0),
\end{eqnarray}
$\{k_i(t)\}$ are called scale-free multiple growing network Markov
chains.
\end{definition}

\begin{lemma}\label{le:3}
If $\lim\limits_{i\rightarrow\infty}d_{k,i}=d_k$ exists and
satisfies $\sum\limits_{k=0}^\infty d_k=1$.
$\lim\limits_{t\rightarrow\infty}tf_t(k):=F(k)$ also exist, and
there is a non-negative integer $m$ to satisfy $d_k=0,
k=0,1,\cdots,m-1, d_m>0$, and $F(k)>0(k=m,m+1,\cdots)$. Then
$\lim\limits_{t\rightarrow\infty}P(m,t)$ exists, moreover
\begin{eqnarray}\label{le:3:1}
P(m)=\frac{d_{m}}{1+F(m)}.
\end{eqnarray}
\end{lemma}
{\bf Proof} With the Markovian property, we have
\begin{eqnarray}
P(m,i,t+1)=P(m,i,t)p_{m,m},\ \ (i\leq t).
\end{eqnarray}

 By the definition of $P(m,t)$ and
$P(m,t+1,t+1)=d_{m,t+1}$, we have the following equation
\begin{eqnarray}
P(m,t+1)&=&\frac{t}{t+1}P(m,t)[1-f_t(m)-\sum_{s=m+2}^{m+m'}
o_{m,s}(\frac{1}{t})]+\frac{1}{t+1}d_{m,t+1}.
\end{eqnarray}
the above difference equation has the following solution

\begin{eqnarray}
P(m,t)&=&\frac{1}{t}\prod_{i=1}^{t-1}[1-f_i(m)-\sum_{s=m+2}^{m+m'}o_{m,s}(\frac{1}{i})]\times\nonumber\\
&
&\left\{P(m,1)+\sum_{u=1}^{t-1}d_{m,u+1}\prod_{j=1}^{u}[1-f_j(m)-\sum_{s=m+2}^{m+m'}o_{m,s}(\frac{1}{j})]^{-1}\right\}.
\end{eqnarray}

Similar to Lemma \ref{le:1} we have Eq (\ref{le:3:1}).

\begin{lemma}\label{le:4}
If the conditions in Lemma \ref{le:3} are all satisfied, and
$\lim\limits_{t\rightarrow\infty}P(k-1,t)$ exists, then
$\lim\limits_{t\rightarrow\infty}P(k,t)$ exists, moreover
\begin{eqnarray}\label{le:4:1}
P(k)=\frac{F(k-1)}{1+F(k)}P(k-1)+\frac{d_{k}}{1+F(k)}.
\end{eqnarray}
\end{lemma}
{\bf Proof} With the property of the Markov chains, we have
\begin{eqnarray}
P(k,i,t+1)=P(k,i,t)p_{k,k}+P(k-1,i,t)p_{k-1,k}+\sum_{l=2}^{m'}P(k-l,i,t)p_{k-l,k}.
\end{eqnarray}

 By the definition of $P(k,t)$ and
$P(k,t+1,t+1)=d_{k,t+1}$, we have the following relation
\begin{eqnarray}
P(k,t+1)&=&\frac{t}{t+1}P(k,t)[1-f_t(k)-\sum_{s=k+2}^{k+m'} o_{k,s}(\frac{1}{t})]+\frac{t}{t+1}P(k-1,t)f_t(k-1)\nonumber\\
&+&\sum_{l=2}^m\frac{t}{t+1}P(k-l,t)o_{k-l,k}(\frac{1}{t})+\frac{1}{t+1}d_{k,t+1}.
\end{eqnarray}

Solve the above difference equation we have
\begin{eqnarray}
&&P(k,t)=\frac{1}{t}\prod_{i=1}^{t-1}[1-f_i(k)-\sum_{s=k+2}^{k+m'} o_{k,s}(\frac{1}{i})]\{P(k,1)+\sum_{u=1}^{t-1}[uP(k-1,u)f_u(k-1)\nonumber\\
&&+\sum_{l=2}^{m'}uP(k-l,u)o_{k-l,k}(\frac{1}{u})+d_{k,u+1}]\prod_{j=1}^{u}[1-f_j(k)-\sum_{s=k+2}^{k+m'}
o_{k,s}(\frac{1}{j})]^{-1}\}.
\end{eqnarray}

Similar to Lemma \ref{le:1} we have
\begin{eqnarray}
P(k)&=&\lim_{t\rightarrow\infty}P(k,t)\nonumber\\
&=&\lim_{t\rightarrow\infty}\frac{tP(k-1,t)f_t(k-1)+\sum_{l=2}^{m'}tP(k-l,t)o_{k-l,k}(\frac{1}{t})+d_{k,t+1}}{(t+1)-t[1-f_t(k)-\sum_{s=k+2}^{k+m'} o_{k,s}(\frac{1}{t})]}\nonumber\\
&=&\lim_{t\rightarrow\infty}\frac{tP(k-1,t)f_t(k-1)+d_{k,t+1}}{1+tf_t(k)+t\sum_{s=k+2}^{k+m'} o_{k,s}(\frac{1}{t})}\nonumber\\
&+&\lim_{t\rightarrow\infty}\frac{\sum_{l=2}^{m'}tP(k-l,t)o_{k-l,k}(\frac{1}{t})}{1+tf_t(k)+t\sum_{s=k+2}^{k+m'} o_{k,s}(\frac{1}{t})}\nonumber\\
&=&\frac{F(k-1)}{1+F(k)}P(k-1)+\frac{d_{k}}{1+F(k)}.
\end{eqnarray}
then complete the proof.
\begin{theorem}\label{th:3}
If the conditions in Lemma \ref{le:3} are all satisfied, then the
steady degree distribution of $\{k_i(t)\}$ exists. Moreover,
\begin{eqnarray}\label{th:3:1}
P(k)=\left\{\begin{array}{ll}
\frac{d_{m}}{1+F(m)}, &\textrm{$k=m,$}\\
\prod\limits_{i=m}^{k-1}\frac{F(i)}{1+F(i+1)}[\frac{d_{m}}{1+F(m)}+\sum\limits_{l=m}^{k-1}\frac{\frac{d_{l+1}}{1+F(l+1)}}{\prod\limits_{j=m}^{l}\frac{F(j)}{1+F(j+1)}}], &\textrm{$k>m.$}\\
\end{array}\right.
\end{eqnarray}
\end{theorem}
{\bf Proof}From Lemma \ref{le:3} and Lemma \ref{le:4}, Eq
(\ref{th:3:1}) comes into existence.

\begin{theorem}\label{th:4}
Theorem \ref{th:2} is also held when allowing multiple linking and
loops.
\end{theorem}

\begin{theorem}
 If $f_t(k)=a_tk+b_t+o_{k}(\frac{1}{t})$, we have $\lim\limits_{t\rightarrow\infty}tf_t(k):= F(k)=Ak+B$ if and
 only if
 $\lim\limits_{t\rightarrow\infty}ta_t=A,\lim\limits_{t\rightarrow\infty}tb_t=B$.
\end{theorem}

\paragraph{Example 1.8}  From the DMS model we get:
$d_0=1,f_t(k)=m\frac{k+H}{(m+H)t}$, so
$A=\frac{m}{m+H},B=\frac{mH}{m+H}$, with Theorem \ref{th:3} and
Theorem \ref{th:4}, we have
\begin{eqnarray}
P(k)=\left\{\begin{array}{ll}
\frac{m+H}{m+H+mH}, &\textrm{$k=0,$}\\
\frac{\Gamma(k-1+H)}{\Gamma(k+1+H+\frac{H}{m})}\frac{\Gamma(H+2+\frac{H}{m})}{\Gamma(H)}\frac{m+H}{m+H+mH}\sim k^{-(2+\frac{H}{m})}, &\textrm{$k>0.$}\\
\end{array}\right.
\end{eqnarray}
So the network is scale-free and degree exponent is $2+\frac{H}{m}$.

\paragraph{Example 1.9} From the model we get
$d_m=1,f_t(k)=m\frac{k}{2mt+m}$,
 so $A=\frac{1}{2},B=0$, with Theorem \ref{th:3}
 and Theorem \ref{th:4}, we have
\begin{eqnarray}
 P(k)=\left\{\begin{array}{ll}
\frac{2}{m+2}, &\textrm{$k=m,$}\\
\frac{\Gamma(k)}{\Gamma(k+3)}\frac{\Gamma(m+3)}{\Gamma(m)}\frac{2}{m+2}\sim 2m^2k^{-3}, &\textrm{$k>m.$}\\
\end{array}\right.
\end{eqnarray}
it is the same as in \cite{b}.

Finally, the authors thank Professor Yirong Liu for his many helpful
discussions. This research was supported by the National Natural
Science Foundation under Grant No. 10671212, and Research Fund for
the Doctoral Program of Higher Education of China No.20050533036.

%
%

%
\end{document}